# Rodlike Complexes of a Polyelectrolyte (Hyaluronan) and a Protein (Lysozyme) observed by SANS.

*Second revision (improvement of English) + SI DESCRIPTION  INCLUDED, and  with Scheme 2 reviewed.*

**I. Morfin[1], E. Buhler[2], F. Cousin[3], I Grillo[4], F. Boué[3]**

[1] Laboratoire de Spectrométrie Physique, CNRS UMR 5588, Université Joseph Fourrier, BP 87, 38042 Grenoble Cedex 9, France

[2] Matière et Systèmes Complexes, UMR CNRS 7057, Université Paris 7-Denis Diderot, Bâtiment Condorcet, CC 7056, 75205 Paris Cedex 13, France

[3] Laboratoire Léon Brillouin, UMR 12 CEA-CNRS, CEA Saclay, 91191 Gif-sur-Yvette, France

[4] Institut Laue Langevin, Large Scale Structures Group, 6 rue Jules Horowitz, BP 156, 38042 Grenoble Cedex 9, France

ABSTRACT

We study by Small Angle Neutron Scattering (SANS) the structure of Hyaluronan –Lysozyme

complexes. Hyaluronan (HA) is a polysaccharide of 9 nm intrinsic persistence length that bears one negative charge per disaccharide monomer ($M_{mol}$ = 401.3 g/mol); two molecular weights, $M_w$ = 6000 and 500 000 Da were used. The pH was adjusted at 4.7 and 7.4 so that lysozyme has a global charge of +10 and + 8 respectively. The lysozyme concentration was varied from 3 to 40 g/L, at constant HA concentration (10 g/L). At low protein concentration, samples are monophasic and SANS experiments reveal only fluctuations of concentration although, at high protein concentration, clusters are observed by SANS in the dense phase of the diphasic samples. In between, close to the onset of the phase separation, a distinct original scattering is observed. It is characteristic of a rod-like shape, which could characterize "single" complexes involving one or a few polymer chains. For the large molecular weight (500 000) the rodlike rigid domains extend to much larger length scale than the persistence length of the HA chain alone in solution and the range of the SANS investigation. They can be described as a necklace of proteins attached along a backbone of diameter one or a few HA chains. For the short chains ($M_w \sim 6000$), the rod length of the complexes is close to the chain contour length ($\sim$ 15 nm).

## I. Introduction

Hyaluronan (HA) is a linear semi-rigid polyelectrolyte with the repeating disaccharide structure poly(($1\rightarrow3$)-β-D-GlcNAc-($1\rightarrow4$)-β-D-GlcA), with global formula $C_{14}H_{20}NO_{11}Na$, for the Sodium salt form (see extended formula of the disaccharide unit in Scheme 1), molar mass 401.3 g/mol and length $b_{segm}$ close to 1 nm. HA is found in large quantities in animals and human body, where it is mainly produced by fibroblasts and other specialized connective tissues cells. It is a common component of synovial fluid and extracellular matrix and one of its assumed functions in the body is joint lubrication.

**Hyaluronan and hyaluronan-protein complexes.** In joints as well in other situations, HA seems to have complex roles and its associations and interactions with proteins or other biomolecules are known to participate in many biological mechanisms that can even be opposite depending on the molecular mass.[1] Large masses are involved in ovulation, embryogenesis, regeneration, while small masses are inflammatory[2], and very short ones are involved in the mobility of cancerous cells and stimulate neo-vascularization of tumors. In pharmacology, HA is widely used in dermatology, ophthalmology, and rheumatology[3]. In this last domain, different formulations of hyaluronic acid are used for arthritic patients, while proof of their efficiency remains to be established[4].

Among the group of glycoaminoglycans (GAGs), HA is the only one that is not covalently associated with a core protein. Being charged, like DNA, HA has specific interactions with proteins (HAbp, HA bonding proteins, like CD44 which is important in cancer for example). In addition, HA has a repetitive primary structure that allows also nonspecific binding with proteins, via electrostatic interactions identical along the chain[5], contrary to DNA. The charge density of the HA depends on pH of the solution since the dissociation coefficient of the carboxyl groups is pH-dependent; such complexes depend therefore on pH (as well as on ionic strength). Detailed analysis and theory has been given by Cleland[6]. The first HA-protein complex structures have been studied by biologists for medical applications, using biological indirect investigation techniques. Only few recent works have investigated HA-protein complexes from a soft matter point of view[5-7]. For interaction of HA with BSA, identification by Integrated Computer Modeling and Light Scattering Studies of an Electrostatic Binding Site has been reported recently[7]. First, the dependence on salt concentration has been studied by Moss[5], while the effect of pH is reported by Malay et al.[8] More recently, the effect of charge ratio has been considered by Lenormand[9]. Bovine serum albumin (BSA) was used as a model to study the HA-protein electrostatic complexes at pH=4 using zetametry and Uronic Assay. At low ionic strength, there is formation of (i) neutral insoluble complexes at the phase separation and (ii) small positively-charged or large negatively-charged soluble complexes whether BSA or HA is in excess. Moreover, hyaluronidase, which catalyses HA hydrolysis, was also shown to form various types of complex with HA at low ionic strength, and to be in competition with BSA. These studies indicate that electrostatic parameters are important. On the reverse, some studies are missing: the effect of H-bonding, largely present between HA chains in pure HA aqueous solutions[10], or the effect of molecular weight, observed in biological situations, have not been investigated. But also, none of these works gives information on the internal structure of complexes.

**Parameters involved in the electrostatic complexes**: Let us first review rapidly the experimental and theoretical works done on polyelectrolytes-proteins complexes or polyelectrolytes-polyelectrolytes complexes during the last few decades in order to detail all relevant parameters for complexation. Soft matter scientists were first interested on synthetic polyelectrolyte-protein complexes for fundamental research or technology. Using essentially techniques leading to general information (turbidimetry, rheology, spectrometry, light scattering or microscopy), studies focused on the solubility of the complexes and established phase diagrams, as reviewed by Tribet[11] and Cooper[12]. Natural polymer-protein complexes studies, taking into account many parameters like pH, ionic strength, and concentrations, have been also recently reviewed by Turgeon et al[13]. Electrostatic interactions have a dominant role in these systems. Charge densities, as shown experimentally, drive all signatures of complexation to their utmost for charge ratio introduced in solution ($[-]/[+]_{intro}$) equal to 1, which corresponds to neutralization of the complexes[14,15] (where the charge ratio $[-]/[+]_{intro}$ corresponds to the ratio between the negative and the positive charges carried respectively by the polymer and the protein). Surface charge densities are heterogeneous when proteins are involved instead of homogeneously charged spheres, so that patches of opposite charges enable electrostatic complexation with polyelectrolyte of the same global charge: this has been observed for polyanions at pH higher than the pI of the protein[16, 17]. Another key factor is the gain of entropy, observed by calorimetry[18, 14, 19], and attributed to the release of the counterions initially condensed on the chains or on the spheres. This has been shown recently for large globular complexes by direct structural measurement (scattering) of the counterion location[20]. However, complex coacervation could be enthalpy driven for less charged systems[21-23].

About ionic strength effect, while some experimental studies show that the Debye length $\kappa^{-1}$ is the determining scale[12], in accordance with simulations, another experiment has shown recently that a decrease of $\kappa^{-1}$ (i.e. an increase of electrostatic screening), can increase the size of protein/ polyelectrolyte globules when they are present in the complexes[24]. The asymmetry of electrostatics around strongly patched proteins can also be important[26].

Nevertheless, interactions other than electrostatics have to be considered, especially in natural polymer or proteins: hydrogen bonding, e.g. for PolyEthyleneOxyde (PEO) - or polysaccharide - based backbones, and hydrophobic interactions if the backbone of the polyelectrolyte is partially hydrophobic, e.g. due to alkyl chains[11, 27]. In the latter case, interactions with hydrophobic patches on the protein can contribute to the formation of a complex[14, 28]. Another effect of interactions can lead to changes in structure of one of the component: this has been observed for sodium polystyrene sulfonate, PSSNa, which unfolds lysozyme and napin protein, but not with another polyelectrolyte[29].

An important parameter, which is relevant to this study, is the rigidity of the polymer chain (characterized by its persistence length $L_p$). All numerical studies underline the importance of the polyelectrolyte stiffness: complexation is abated when the polyelectrolyte is less flexible[30, 31]. These main trends have been checked experimentally by Kayitmazer et al. who compare the binding of chitosan and PDADMAC, two polyelectrolytes with equal charge density but different persistence lengths, to oppositely charges micelles, dendrimers and proteins[32]. However the authors point out that the relevant parameter for binding is not necessarily the intrinsic persistence length of the chain $L_p$, but the flexibility of the chain on the colloid length scale[32].

In parallel, computer simulations also studied the role of electrostatic interactions in mixtures of charged strings with charged spheres and their interplay with parameters such as chain stiffness, at first in simulations involving only one protein and one polyelectrolyte chain[30, 31], and more recently on systems dealing with several proteins in the presence of an oppositely charged polyelectrolyte[33, 34]. In ref [34], the heterogeneous charge density of proteins in the presence of weak polyelectrolyte has also been taken into account, whereby different chain lengths and ionic strengths were employed:

- large polyelectrolyte-protein clusters are simulated for the strongest electrostatic interactions, at least at stoichiometric charge ratio (with polyelectrolyte in excess, re-dissolution occurs, with proteins still attached to polyelectrolyte).

- clusters are smaller with shorter chains, and for increased ionic strength.

- complexation is reduced for more rigid chains.

Analytical theory has also been used. Reference [35] considers asymmetric mixtures of large macroions – sphere or chain, with smaller ionic species (called Z ions) of opposite charge, which can also be a short chain or a small sphere or ion. The different types of complexes are determined by the electrostatic screening and by the ratio [-]/[+]$_{intro}$ of total charges of Polycation and Polyanion in the solution. When this ratio [-]/[+]$_{intro}$ ~ 1, the authors[35] predict a condensation, that is the formation of "condensates" containing many large components (the macroions) – and therefore also many small Z-ions - with a spherical dense shape (which could be also called a globular shape). This regime of total condensation is sandwiched (for [-]/[+]$_{intro}$ < 1 or >1) by a regime of mixed complexation, where the condensates coexist with smaller complexes which could contain only one large macroions (and several small Z-ions) and could then be called "single complexes". In the case of small spherical Z-ions plus long macroions chains (flexible or semi-flexible), these single complexes are of the bead-on-a-string type, also called by authors "artificial chromatin". Even further from the point [-]/[+]$_{intro}$ ~ 1, only small complexes are present.

**Experiments on internal structures of complexes:** We wish to recall here some of the works where the internal structure, i.e. at distances larger than the protein size and lower than the cluster size, as well as the composition of complexes, has been described accurately using Small Angle Scattering techniques. This is the case of the work performed on polystyrene sulfonate-lysozyme complex systems using Small Angle Neutron Scattering (SANS) [15, 24, 29], for which the systematic use of contrast matching enabled the investigation of the internal structures of the different compounds separately inside the complexes. [-]/[+]$_{intro}$ was established as a key parameter among several different general types of structures. For all other parameters fixed, as [-]/[+]$_{intro}$ varies, both species are quasi-totally associated in one of the following structures: (i) a gel where polymer chains are crosslinked by the proteins when polymer is in excess[29] or (ii) clusters with a well defined spherical shape, rather dense, which we called "globules": inside the two species are in close contact when proteins are in excess[29, 15]. The globules are organized in fractal bunches as confirmed in real space by Electronic Microscopy[15]. SANS was also used recently on complexes of lysozyme and pectin, which being a natural polysaccharide could not be deuterated, but being semi-rigid ($L_p \sim 8$ nm) allowed to check the effect of chain stiffness[36]: globules are also present here, and seem to be less dense (e.g. twice less dense) for such less flexible pectin chains. In spite of the complexity of pectin structure, electrostatic interactions, through the linear charge density, appear as the main parameter in the formation of large dense globules: hydrophobicity has a minor role. SANS (combined with SALS) has also been used to study coacervates of BSA with chitosan[37] ($L_p \sim 60$ nm), which are largely solvated (84% of water), while they are denser ($\sim 75\%$ of water) with the same protein and a more flexible polyelectrolyte, PDAMAC ($L_p \sim 2.5$ nm) [38], which points out again a role of polyelectrolyte chain stiffness. More recently, SANS was used for β-lactoglobulin/pectin coacervates, and showed two kinds of microstructures, either protein molecules dispersed in the coacervate network, or protein domains enclosed in the interpenetrated pectin chains[39]. Networks of polymer-rich zones were also found for agar-gelatin complex coacervates, but with a correlation length around 1.2 nm only[40].

The goal of the present work is the investigation of HA-lysozyme complexes for different HA molecular weights, pHs, ionic strengths of the solution, as a function of the charge ratio between the polymer and the protein. The choice of the HA polymer first comes from its primordial role in biological process as mentioned above but also because it is an excellent model of semi-rigid polyelectrolyte. Indeed, in spite of its natural origin, HA is a perfect linear homopolymer with a well defined intrinsic persistence length, i.e. in absence of electrostatic repulsion, of $L_{p0} \sim 80$ nm. This is well documented since HA has been studied as single specie in aqueous solution for some time[41]. We took particular advantage of direct studies of conformation using also neutron (combined with light) scattering, in pure solution[42] as well as in a mixture with another polyion, sodium polystyrene sulfonate (PSSNa) [43]. The choice of the lysozyme as protein is twofold as well: we have a precise knowledge of the local structure of complexes of lysozyme with other polyelectrolytes, either flexible in case of PSSNa-lysozyme system[15, 29] (where the flexible polyion conformation was measured even inside the complexes[25]), or rigid in pectin-lysozyme system[36], both studied by SANS. In addition, lysozyme is, like HA, present in the extracellular medium of the cartilage, whereas the roles of this protein, or the protein-HA complexes, have not yet been precisely established. This system has therefore also a biological interest. Moreover, extensions to mixtures of HA with other proteins are foreseen.

## II. Materials and methods.

### II.1. Components of the complexes and mixing procedure.

Two very different molecular weights of sodium hyaluronan were used: HA1, of small mass (Mn = 4700, $M_w$ = 6200, Soliance, Pomacle, France), and HA2, of large mass ($M_w$ > 500000, Sigma-Aldrich, ref. H9390). Knowing the segment mass (401.3 g/mol) and its length ($b_{segm}$ = 1 nm) the contour lengths are respectively $L_c$ = 15 nm and 1250 nm. It was mixed with hen egg white lysozyme (HEWL, Sigma Aldrich, ref. 62970), well known as globular, enveloped inside an ellipsoid of size 4.5 x 3 x 3 nm$^3$. Complexes were prepared at two different pHs in order to change $[-]/[+]_{intro}$ keeping the same concentration of the compounds. First pH = 4.7, used before[15, 29], was chosen because it is easily reached in acetic acid/sodium acetate buffer (0.05M/0.05M), and it is a good compromise for both the charge of lysozyme and HA, which remains still high. The ionic strength of this buffer 1 is $I_{buffer}$ = 0.1 M. Buffer 2 was prepared at pH = 7.4, in order to go towards pH conditions in tissues rich in HA (Extra Cellular Matrix, ECM). We used the following Phosphate Buffer Solution ("PBS") contents: to 10 mM of $Na_2HPO_4$ where added 1.76 mM of $KH_2PO_4$ and 2.7 mM of KCl. Following this composition and the literature[44], the ionic strength $I_{buffer}$ is around 30 mM. To be closer to physiological conditions of ECM, to the mixture of such composition we added 137 mM of NaCl, leading to buffer 3, where total ionic strength $I_{buffer}$ is 170 mM. Samples made with this buffer are called "with salt".

All samples for SANS have been prepared the same way. Solutions of HA at 20 g/L (0.5 ml) and lysozyme at 40 g/L (or 80 g/L for few samples) were first prepared in the corresponding buffer 1, 2 or 3, made with $D_2O$. Complexes were formed by adding various volumes of lysozyme and buffer solutions to HA solutions in order to have a final HA concentration equal to 10 g/L. Mixing is first done with a vortex. Temporary precipitation takes place instantaneously as lysozyme is added in all samples. After slow magnet stirring of the samples for at least 24 hours, they reach a stable state that is maintained a few weeks. Such protocol has been chosen as the best after that several others were investigated. SANS measurements are performed a few days after the mixture preparation, when solutions do not present any evolution with time, and before degradation of HA or lysozyme. In any cases, mixing the samples for a longer time does not change the result. The other ways of mixing investigated before this one seems to show the same results after 24 or more hours. We performed SANS measurements just after that the visual aspect became stable, i.e. after 3 days and compared them to measurements after one week: they give the same SANS profiles. Visual observation of the samples a couple of weeks after the SANS measurements neither reveals any modification. We therefore assumed that samples are stable or at least have a time evolution longer than 2 weeks.

We define the charge ratio $[-]/[+]_{intro}$ as the ratio between negative charges of hyaluronan at this pH, and the global charge of lysozyme at the same pH. Here $[-]/[+]_{intro}$ will decrease while lysozyme is added at constant concentration of HA, 10 g/L. At both pHs, HA is quasi-completely or completely dissociated[9], and the number of charges is around 1 charge each monomer unit (which comprises 2 oses, see Scheme 1). For $C_{HA} = 10$ g/L solutions, this gives $10/401 = 0.025$ negative charge per liter (and therefore an additional ionic strength due to sodium counterions of 12.5 mM). At pH = 4.7, lysozyme (molar mass 14298) carries 10 charges per molecules, hence a positive charge molarity ($(10/14298)*C_{Lyso}$ moles/L of charges). The counterions, assumed monovalent, bring an additional ionic strength of 0.33 mM per g/L (e.g. 1 mM at 3.32 g/L and 13.3 mM at 40 g/L). At pH = 7.4, lysozyme has still a global charge positive but 20% smaller ($(8/14298)* C_{Lyso}$).

**Table 1** and **Table 2** list the sample concentrations, and $[-]/[+]_{intro}$ for the two pHs for the large molecular weight HA2. **Table 3** gives the characteristics of the different mixtures for the small molecular weight polymer, HA1, which was only studied at pH 7.4.

At the concentration 10 g/L, HA is in semidilute regime for the large mass 500 000: assuming a fully stretched chain of length $L_c$, the radius of gyration $R_g = L_c/\sqrt{12} = 360$ nm and the overlap concentration is very small, $C^* = M/(4\pi R_g^3/3) = 4.25 \cdot 10^{-3}$ g/L. Conversely, solutions of small mass HA1 are below c* (M = 5000, $L_c = 15$ nm, $C^* = 25$ g/L).

### II.2. SANS measurements.

SANS measurements were performed on PACE spectrometer (Orphée reactor, LLB, CEA, Saclay, France) and on D11 and D22 spectrometer (ILL, Grenoble, France). **Table 4** summarizes the different configurations (wavelengths $\lambda$, sample-detector distances, collimations...) leading to the following q range $3.5 \ 10^{-3} < q \ (Å^{-1}) < 3.5 \ 10^{-1}$, where the wave vector, q, is defined as $q = 4 \ \pi/\lambda$ $\sin(\theta/2)$, $\theta$ being the scattering angle. Each time, detector efficiency was accounted by dividing by the scattering of water ($H_2O$), an almost pure incoherent scatterer, which was also used as absolute intensity standard. All data are therefore in $cm^{-1}$.

We give below in **Table 5** the scattering length density of the different species. First, for HA, $\rho_{HA}$ is calculated assuming complete dissociation of the sodium ion. This dissociation into water has a negative contribution of a few $Å^3$ per ion (-6.6 $cm^3$/mole of ion, i.e. for a solution at 1 Mol/L, a relative variation of $-7.10^{-3}$) distributed over all the solvent volume, as explained in former papers[45]. In terms of contrast this is equivalent, keeping the solvent density constant, to a relative increase of the volume of the HA unit $V_{molecular}$ by $+ 7.10^{-3}$, giving a quantity of the order of $+ 5 \ Å^3$, which is minute. Second, more important, is the effect of exchange of labile protons with the deuterons of the solvent. They are 5 per repetition unit, so that in a $H_2O/D_2O$ mixture with a volume fraction of $D_2O$ $x_{D2O}$, $\rho_{HA}$ becomes $\rho_{HA} (x_{D20}) = 2.3 + x_{D20} (5(b_D-b_H))/V_{molecular}$ (see **Table 5** for 100% $D_2O$). We see also from **Table 5** that the matching point of scattering of the two species HA or lysozyme in a proper $H_2O/D_2O$ mixture would result in relatively small contrast for the other. We thus did not attempt matching experiments.

## III. Results

### III. 1 Macroscopic aspects.

The macroscopic visual aspect of the different samples is summarized in **Scheme 2**. In all cases without salt, below a particular $[-]/[+]_{intro}$ that can be slightly different for a sample to another, a phase separation is observed, between a supernatant solution (clear or turbid liquid), and a lower, thus denser, turbid one at the bottom of the cell. This lower phase looks like a gel for the larger polymer mass, and like a precipitate for the smaller one (in this case it that can be re-suspended by agitation for the time of measurement).

Scheme 2 must not be considered as a phase diagram. We will not discuss in details the reason of the slight [-]/[+]$_{intro}$ difference for a same event between the two pHs (which are not at the same ionic strength). The goal of the paper is not to investigate in detail the effect of pH or of the ionic strength, but rather to overview the general behavior of HA-lysozyme mixtures when [-]/[+]$_{intro}$ is varied. We will then see in the following whether the similarity between samples observed macroscopically is also found for scattering results.

## III. 2. Scattering from single species.

**Hyaluronan scattering:** **Figure 1** shows the scattering of the pure HA2 solution having the largest molecular weight, HA2. This rather weak signal was measured here only for reference. At large q, it displays a $q^{-1}$ variation, within the error in background subtraction. When q decreases, the scattering crosses over, around $5 \cdot 10^{-2}$ Å$^{-1}$, from a $q^{-1}$ to a $q^{-2}$ variation which is characteristic of a flexible chain at a scale larger than the persistence length[46]. Such cross-over value agrees with the value of the persistence length $L_p$ which has been determined in various conditions of polymer and salt concentrations directly on the form factor using SANS for HA[42]. An intrinsic value $L_{p0} > 8$ nm at least was found (a smaller value of 4 nm was given formerly using less direct methods[41]), and due to electrostatic repulsion it can increase for weak ionic strength I as predicted by the Odijk-Skolnick-Fixman model ($L_{pe} = L_p - L_{p0} \sim \Gamma^{-1}$)[47, 48]. Here the total ionic strength is I ~ 50 mM, making $L_{pe}$ a few Angströms, which brings $L_p$ close to $L_{p0}$. The most important for the following is probably that the HA signal is always rather weak, compared to the signal measured for all samples except at large and low q for the complexes with $C_{Lyso} = 3.32$ g/L. The low q upturn is due to the well known presence of some HA aggregates; its influence is discussed below. SANS measurements of pure solutions of HA1 (lowest molecular weight), not shown here, show also a usual behavior.

**[Figure 1]**

**Lysozyme scattering:** In the same **Figure 1**, the top curve is the scattering from a pure lysozyme solution, at 20 g/L in buffer at pH = 7.4 (I = 0.01 M):

- at large q, above 0.2 Å$^{-1}$, it varies like q$^{-4}$ (Porod scattering associated to the sharp interface between the compact globular protein and the surrounding medium).

- between 0.2 Å$^{-1}$ and 0.01 Å$^{-1}$, the curve flattens: this corresponds to the Guinier regime q. $R_{gLyso}$ < 1, ($R_{gLyso}$ radius of gyration of the protein); it must tend to a constant proportional to the mass and concentration of the protein.

- however, in the lowest q points, an upturn appears. Part of the increase of the zero-q limit could be associated with dimerisation – j-merisation of lysozyme[49], but the main effect is likely due to larger protein aggregates[50]. Neglecting this slight upturn, the signal can be extrapolated to a zero q limit of 0.28 cm$^{-1}$, in agreement with contrast, volume, and density of lysozyme given in Table 5 [NOTE 1]. It can also be fitted to the form factor of a polydisperse sphere of radius R = 1.55 nm as shown on **Figure 1** with a slightly lower zero q limit, 0.23 cm$^{-1}$. We consider this difference negligible for our goal. The goal is to use this spectrum to compare with the HA signal. Assuming the signal as proportional to concentration at lower concentration, we extrapolate the signal at 3.32 g/L from the signal at 20 g/L, by dividing it by the concentration ratio 20/3.32, as also shown in **Figure 1**. This corresponds to the lowest lysozyme concentration used for the complexes. We see that even at this concentration, the protein scattering is larger than the HA signal in almost all q regions except at q < 0.014 Å$^{-1}$, and also close to the largest q values. At larger $C_{Lyso}$ the scattering of the protein is larger than the one of the HA for all q values. This remark prompts us to consider the HA signal to be a negligible contribution for lysozyme concentrations larger than 3.32 g/L. We will discuss the latter case in detail below.

### III.3. Scattering from large molecular weight HA2-lysozyme complexes with low salt buffer.

### III. 3. 1 Overview.

**Figure 2 and 3** are the SANS curves of different HA-lysozyme mixtures respectively at pH = 4.7 and 7.4. All scattering intensity data of figure 2 are cm$^{-1}$, but each curve, when going from the bottom to the top, is shifted from the former one by a factor 10 in the log scale. The bottom one ([-]/[+]$_{intro}$ = 10.7) as well as the signal of the pure HA solution shown as reference are in real units (cm$^{-1}$, not shifted). For the two smallest charge ratios (top of the Figure, area labelled in grey), are plotted the SANS profiles of the dense phase. For the other monophasic samples and for samples for which the dense phase was in too small amount (15 and 20 g/L), the scattering comes from the liquid phase. In **Figure 3**, the intensity shifts correspond to the same multiplication factor for the same lysozyme concentration than on **Figure 2**.

**[Figure 2]**

**[Figure 3]**

For both pHs, neutron scattering from the different mixtures signals the arising of new structures, different from a simple mixing of the two components. Such structures vary largely with [-]/[+]$_{intro}$ and the scattering profiles depend largely on the scattering vector q. Although [-]/[+]$_{intro}$ are sometimes different from one pH to the other for a same event, some similarities appear in the two figures where three regimes of [-]/[+]$_{intro}$ can be determined:

- at low protein fractions (high [-]/[+]$_{intro}$), where an upturn that can not be attributed to the scattering of pure lysozyme and HA solutions is observed at low q,

- at intermediate ones, where the scattering is characterized by a well defined q$^{-1}$ region at

intermediate q,

- at large protein fractions (low [-]/[+]$_{intro}$) where the previous q$^{-1}$ power law disappears.

In the following, some of the scattering curves of these two figures will be selected and redrawn in complementary figures in order to discuss the structures at the different lengths scales and for the different charge ratio regimes.

**III. 3. 2. Detailed SANS analysis.**

**III.3.2.1 Low protein fractions.**

We give details here on the first type of profile, at high [-]/[+]$_{intro}$, taking **Figure 4** for illustration (pH 7.4). Here we present the scattering of a lysozyme corresponding to a 3.32 g/L solution. Compared with this scattering are three samples (shown already in **Figure 3**) having the three lowest protein contents (3.32 g/L and 6.64 g/L and 8g/L of lysozyme, [-]/[+]$_{intro}$ = 13.4, 6.7 and 5.5). The third one (5.5) being diphasic, only the liquid fraction has been observed.

[Figure 4]

At large q, above 0.2 Å$^{-1}$, the scattering varies as a q$^{-4}$ law, characteristic of sharp interfaces. **Figure 4** shows that it is possible to superimpose the data to the lysozyme alone at 3.32g/L by multiplying them by a given factor, *F*. For monophasic mixtures with 3.32g/L and 6.64g/L of lysozyme, *F* corresponds to 1 and ½, respectively: the high q signal corresponds to the lysozyme form factor. This also demonstrates that the proteins keep their native shape (no unfolding) and are still surrounded by solvent (their interface with solvent is still visible). The intensity of the signal is also the sign that the sample is monophasic. On the contrary, for the mixture containing originally 8g/L of lysozyme, the factor is ½ instead of 1/2.4, (from the ratio of the nominal lysozyme concentrations: 8/3.32) because the sample is diphasic: the missing lysozyme is indeed in the dense phase, which has sedimented at the bottom of the cell. This remark is true for all scattering curves of diphasic samples shown in this work.

At intermediate and small q values, the sample scattering flattens with q decreasing, (as was observed on the pure lysozyme scattering (see III.2); this corresponds to a Guinier regime) but this flattening is followed at even lower q by an upturn for the three concentrations between 3.32 g/L and 8 g/L, as gathered on **Figure 4** for pH 7.4. The 3 data for pH 4.7 also show this upturn in **Figure 2**. This upturn *cannot* be attributed to the HA scattering contribution: this is established through **Figure S.I.2** of Supporting Information: the scattering of the complexes for $C_{Lyso} = 3.32g/L$ (the lowest concentration) is plotted and compared with the same spectrum after subtraction of the HA contribution. In the latter the upturn stays well defined with a slope close to -2.1 +/- 0.02, indicating that a low q scattering identical to the one of the pure HA solution is *not* the major contribution to the low q scattering of the complexes.

### III.3.2.2 Intermediate protein content and charge ratio (between monophasic and fully diphasic systems): rod scattering.

**Figure 5** shows the scattering of mixtures in the intermediate range of charge ratio ($2.2 \leq [-]/[+]_{intro} \leq 3.6$) at pH = 4.7, shown already in **Figure 2**. The scattering corresponds either to monophasic and turbid samples, or to the liquid turbid part of the first diphasic sample (the dense phase has a small volume located at the bottom of the cell). In this charge ratio range, the scattering keeps overlapping the form factor of the lysozyme at large q, as seen in insert of **Figure 5**, while a new type of variation with q, never seen before in protein-polyelectrolyte complexes at our knowledge, appears at lower q: a straight line corresponding to a $q^{-1}$ power law. Such $q^{-1}$ regime spreads over a large range of q, reaching the smallest q available in our experiments ($3.10^{-3}$ Å$^{-1}$) for a particular charge ratio ($[-]/[+]_{intro} = 3.6$, $C_{Lyso} = 10$ g/L, pH = 4.7) as seen in **Figure 5**. The presence in solution of rigid domains that extend to larger length scale than the persistence length of the HA alone in solution may explain such power law. When varying the lysozyme concentration, the full curve is shifted by the same factor at all q. In other words, both the part of the curve showing the $q^{-1}$ regime, and the part at large q corresponding to the lysozyme form factor, are multiplied by the same factor: this suggests strongly that the scattering is dominated by the lysozyme, and that all the proteins are belonging to the complexes.

[Figure 5]

### III.3.2.3 Diphasic systems and globular objects.

A further increase of the lysozyme concentration leads to diphasic samples with an increasing quantity of dense phase. The SANS $q^{-1}$ variation observed previously disappears at the two q range boundaries:

- at the low q, an upturn grows progressively and extends up to q $\sim$1 $10^{-2}$ $\text{Å}^{-1}$. At pH 4.7, the upturn is first seen –this is moderate for this sample little- for $[-]/[+]_{intro}$ = 2.4 ($C_{Lyso}$ = 15 g/L), where a tiny volume of dense phase is seen, corresponding to the onset of phase separation. The upturn becomes clearly visible for $[-]/[+]_{intro}$ = 2.2 ($C_{Lyso}$ = 20 g/L) just above the onset. For higher protein contents (27 and 40 g/L), the volume of the dense phase increases and the upturn also increases; for 40 g/L it reaches a $q^{-4}$ law (Porod scattering). This second Porod law at low q is the signature of large compact objects, which can contain both lysozyme and HA. This $q^{-4}$ variation extends towards the lowest q values (as can be checked using $q^4 I(q)$ representation in **Figure S.I.2** of Supporting Information): this means that the average global size of these objects is larger than 60 nm. We will call these objects "clusters" in the following.

- at the large q, around $q_{max}$ $\sim$ 0.2 $\text{Å}^{-1}$, for the highest protein contents (27 and 40 g/L), the scattering of the dense phases is dominating since they occupy most of the volume. It has not the same shape than pure lysozyme at large q: a "shoulder" is observed in the scattering. It thus appears when the cluster scattering and the phase separation are sufficiently important. It could be the signature of proteins in close contact inside clusters, at a distance D = 2 $\pi$ / $q_{max}$ $\sim$ 3.1 nm. This distance D is close to (2. $R_{lyso}$), where $R_{lyso}$ is the average lysozyme radius. A correlation peak has been observed formerly in different systems, in particular for lysozyme mixed with PSS or pectin at the same q abscissa[13, 34].

**[Figure 6]**

At pH 7.4 in **Figure 3** a similar behavior than in **Figure 2** is observed**:** an upturn as soon as $C_{Lyso}$ = 8 g/L, and a shoulder as soon as $C_{Lyso}$ =10 g/L. But, moreover, we can compare in **Figure 6**, the scattering measured completely separately for the two phases, liquid supernatant (two lower curves) and dense phase (two upper curves), for $C_{Lyso}$ = 10 and 40 g/L. The behaviors are different. For $C_{Lyso}$ =10 g/L, at smaller q, the signal of supernatant is close to the monophasic solution at 6.6 g/L (also shown on **Figure 6**), while the dense phase shows aggregation and a protein-to-protein shoulder. This indicates that the supernatant phase in equilibrium with the dense phase is close in structure to the monophasic samples. On the contrary, for $C_{Lyso}$ = 40 g/L, the signal of supernatant is the one of single lysozyme, while the dense phase shows even more pronounced aggregation and protein-to-protein shoulder (while the $q^{-1}$ rod signal is still present). In the latter case, the supernatant signal suggests therefore that phase separation is completed: all complexes have precipitated into the dense phase, while the supernatant keeps the lysozyme in excess.

**III.4 Large molecular weight HA2-lysozyme complexes with salt (physiological conditions of extracellular matrix).**

The effect of salt (in PBS buffer) has been investigated for complexes made with high molecular weight HA (HA2). 137 mM of NaCl have been added to the initial buffer solution (final value $I_{buffer}$ = 170 mM) before mixing in the same proportions of HA2 and lysozyme than some of the samples without salt. In presence of salt, whatever the quantity of lysozyme added, samples never show a macroscopic phase separation. Nevertheless, for the largest lysozyme amounts (40 g/L), they are turbid and look like paste.

**[Figure 7]**

**Figure 7** compares the scattering curves from samples with and without salt (pH 7.4). The behaviors are different at the two protein concentrations:

- for [-]/[+]$_{intro}$ giving monophasic samples in low salt conditions, the SANS profiles are not clearly modified by addition of salt.

- for [-]/[+]$_{intro}$ giving diphasic samples in low salt conditions, both the shoulder at large q, attributed to protein-to-protein stacking and the low q upturn disappear when adding salt.

Thus the formation of clusters is prevented by salt, probably due to electrostatic screening, while the single complexes observed in monophasic samples are not salt sensitive, at least at the scales observed by SANS.

**III.5 Small molecular weight HA1-lysozyme complexes: short rods.**

**Figure 8** shows the scattering of the different samples listed in **Table 3** (buffer pH is 7.4, $I_{buffer}$ = 30 mM and total ionic strength varies between 44 mM and 55 mM depending on lysozyme content). The highest concentrations give diphasic samples, but the dense phase is composed of small solid objects visible by eyes which can be put in suspension by shaking the samples. For the lowest [-]/[+]$_{intro}$, ($C_{Lyso}$ = 40 g/l), the scattering has been measured for the liquid supernatant only, as well as on the whole sample, after shaking the sample a few seconds just before recording the data: the two scatterings are essentially very close (**Figure 8**). The scattering at $C_{Lyso}$ = 27 g/L scale is also very close. At large q, a $q^{-4}$ power law is observed, followed at intermediate q by a $q^{-1}$ slope. Comparison with scattering from HA2 is enlightening:

- in the $q^{-1}$ range, the intensity almost overlaps with the one for the large mass (HA2) at same pH

and same concentration, or shows proportionality to concentration [NOTE 2]. This means that the protein linear density is the same along short and long chains.

- but at low q, very interestingly, the scattering is different from the HA2 one: for the small mass it reaches a plateau. Thus the radius of gyration, hence the length of the rodlike complexes, are smaller. Indeed the scattering matches a model of finite length cylinder, as shown in **Figure 8** for $C_{Lyso}$ = 40 g/L (the same fit can be performed at $C_{Lyso}$ =27 g/L). We find a rod length of around 15 nm, close to the average contour length of a short HA1 chain (see II.1). A more accurate fit would require a more accurate knowledge of the molecular weight distribution of HA but would not modify the main conclusion: the rod length is the contour length of one chain. This excludes the fact that different short chains would connect at their ends, i.e. build some kinds of "splices", to form longer rods.

**[Figure 8]**

**IV. Discussion.**

Let us first summarize the experimental facts: for the three systems, where the HA concentration is constant, a consistent picture arises (with many common features at pH 4.7 and 7.4):

(i) At low protein content the low q scattering (recorded for long chains only) presents an important upturn with a well defined $q^{-2.1}$ variation. We have shown that it cannot be attributed to the pure HA solution scattering. Therefore it should come from concentration fluctuations of HA or lysozyme. The fact that this signal is close to proportional to protein concentration at constant HA concentration suggests that it comes from proteins, and can be interpreted by concentrations inhomogeneities. In summary, except for inhomogeneities of concentration seen at low q lysozyme appears dispersed in the network of hyaluronan chains.

(ii) For higher protein content, we observe the central feature of this paper: a $q^{-1}$ power law which can be interpreted, using absolute units, as rod-like alignments of proteins, probably along HA chains. The apparent axial radius of the rod is the average radius of the lysozyme. For long HA chains, the rod length is much larger than the sizes accessible in our q range, thus can be as large as the length of the chains. For much shorter contour length HA chains, the rods are also much shorter, of the order of the chain contour length.

(iii) At protein concentration slightly larger than the threshold for the rod behavior ($C_{Lyso\_rod}$) for pH 7.4, or simultaneously to its onset, for pH 4.7 (lysozyme is 20% more charged), two facts occur for samples without salt:

- macroscopically, the mixtures phase-separate.

- submicrocopic scales are investigated by scattering, which displays, for long chains, an upturn at low q (not investigated for short HA1 chains), overshadowing the rod signal, while a shoulder appears simultaneously at large q. This can be due to the formation of large dense clusters, like found for lysozyme with other polymers. With PSS and with pectin, which both show dense globules, the low q features are similar, and at the q abscissa of the shoulder a peak is present, unambiguously associated to contact between two proteins[27, 34].

Let us now discuss the picture we propose, in particular for the rod like structure, and their relation with phase separation. The fact that the rod scattering is proportional to $C_{Lyso}$ (in monophasic samples) suggests that the number of lysozyme per rod does not change deeply until phase separation occurs. This is seen also in the way the scattering crosses over with the large q single lysozyme scattering: if the protein linear density was increasing, the front factor of the $q^{-1}$ regime would vary, resulting in a different aspect of the connection with the large q scattering. What we see suggests that when increasing $C_{Lyso}$ around $[-]/[+]_{intro} = 5$, more "decorated rods" of same density are formed. When increasing $C_{Lyso}$ a little more, the solution becomes too concentrated in these decorated rods, and starts to separate in two phases. Data of **Figure 6** suggest that some rods may stay present in the supernatant while they can also be observed in the dense phase (see Discussion above). However, soon above the phase separation threshold for $C_{Lyso}$, a low q upturn points out the formation of larger clusters. Here again they can be detected in both phases, but their scattering becomes dominant at the highest lysozyme concentrations, in the dense phase. It is possible, but not proven, that the clusters contain rods. For the largest added amount of protein, all complexes are in the dense phase, where many clusters are present.

Two next questions, actually correlated, concern the composition in lysozyme and hyaluronan of the rods: (i) what is their charge, i.e. are they neutral, and (ii) how many HA chains do they contain? As inferred by the concentration dependence of $I(q) \sim C_{Lyso}$ shown in **Figure 5**, we can neglect the scattering coming from the HA. We can then quantify the dressing effect as follows. Let us consider, in a volume $\mathcal{V}$, $\mathcal{N}_{rods}$ rods of length L. We define $\Delta\rho_{Lyso}^{2}$ the contrast between the lysozyme and the solvent, $\mathbf{v}_{Lyso}$ the molecular volume occupied by one lysozyme, N is the number of lysozyme per rods. For $q \gg 1/R_{grod}$ ($R_{grod} = L/\sqrt{12}$), the expression for a $q^{-1}$ scattering of a rod dressed at random by N proteins is

$$I(q), \text{cm}^{-1} = \Delta\rho_{Lyso}^{2} \cdot (\mathcal{N}_{rods}/\mathcal{V}).(N \cdot \mathbf{v}_{Lyso})^{2}.\pi/q.L \qquad (1a) \qquad q \gg 1/R_{grod}$$

$$= \Delta\rho_{Lyso}^{2} \cdot \Phi_{Lyso}. \mathbf{v}_{Lyso} \ \pi/q.a, \qquad\qquad (1b)$$

defining the protein volume fraction : $\Phi_{Lyso} = (\mathcal{N}_{rods}/\mathcal{V}).$ (N. $\mathbf{v}_{Lyso}$), and $N/L = 1/a$. The volume of protein per length, $\mathbf{v}_{Lyso}/a$, is the relevant parameter given by the $q^{-1}$ law. We use data for 10 g/L of lysozyme because we can safely assume that the nominal concentration is the actual one for this sample; and that all proteins belong to the rods ($I(q) \sim \Phi_{Lyso}$, as discussed above). Fitting the curve in the $q^{-1}$ regime to Eq. (1b), we obtain a value for the front factor of the $1/q$ law, $F(\Phi_{Lyso}) = \Phi_{Lyso}$ . $\Delta\rho_{Lyso}^{2} \cdot \pi \ \mathbf{v}_{Lyso}/a$.

We can then use the fact that the zero q limit of the scattering measured for pure lysozyme is:

$$I_0 = \lim_{q \to 0} I(q) = \Phi_{Lyso} \cdot \Delta\rho_{Lyso}^2 \cdot v_{Lyso} \quad (2)$$

From the values of $F(\Phi_{Lyso})$ and of $\lim_{q \to 0} I(q)$ we get[NOTE 3] a value $a \sim 8$ nm.

To end this part, we note that the contour length of a HA chain of molar mass 500 000 is $L_{cHA}$ = 1250 nm (see II.1). If the rod has the same length, its radius of gyration will be equal to $R_{grod}$ = $L/\sqrt{12} \sim 360$ nm. This makes that $q.R_{grod} \gg 1$ down to our lowest q accessible: the Guinier regime lies at much lower q than available, and a more detailed mathematical fit is not useful.

Our result suggests that proteins are aligned, with a linear density of one protein (of size $4.5 \times 3.0 \times 3.0$ nm$^3$) over a length $a = 8$ nm (+/- 10%.). Assuming $\mathcal{N}_{rods} = \mathcal{N}_{HA}$, i.e. one single HA chain in the core of the rod, with one monomer each 1 nm, this would give a ratio of 1 lysozyme for 8 HA segments. Since the global lysozyme charge is of + 10 at pH 4.7, the rodlike complex would be close to neutral, i.e the inner charge ratio within a rod $[-]/[+]_{inner} \sim 1$. This is far from the average introduced charge ratio, here $[-]/[+]_{intro} = 3.6$ for $C_{HA}$ = 10 g/L, and $C_{Lyso}$ = 10 g/L, which corresponds to a ratio of 1 protein for 35 HA segments. Note that it is possible that some chains pertain to dressed rods, in coexistence with others which are free. In this case, $[-]/[+]_{inner}$ could be equal to 1 over a large lysozyme concentration. This would likely account for phase separation of neutral non repulsive species over a wide range of $[-]/[+]_{intro}$. New neutral rods are progressively formed when protein is added, and progressively separate into the dense phase. When $[-]/[+]_{intro}$ approaches 1, all HA has been consumed, and we obtain the situation where all complexes are in the dense phase in the bottom of the cell (see results of **Figure 6** for 40 g/L of lysozyme). This scenario involves "disproportionation" between neutral complexes and free charged species.

An alternative is that the number of HA per rod $N_{HA}/N_{rods}$ is larger than 1: this agrees with data since Eq.(1b) is not sensitive to $N_{HA}/N_{rods}$. Values slightly larger 1 should not affect the scattering since the scattering of individual HA chains is around 20 times lower than the rod one (see insert of **Figure 5**). A ratio of 1 protein for 35 HA on each rod, corresponding to the hypothesis that there are no free species in solutions, could thus be reached with $35/8 \sim 4.5$ chains per rod, making them strongly charged because one would obtain $[-]/[+]_{inner} = 3.6$ in this case. It is then surprising that phase separation occurs at concentration just above 10 g/L. We can also remark that bundles of several HA per rod makes possible to create rods longer than one HA chain contour length, by connecting chains along themselves ("splices"). This is not supported by observations in the case where we could check this in practice, namely for short HA (HA1): the rod length given by the fit is equal to the contour length of HA extracted from the molecular weight within our accuracy. All together we do not favor such "multi- HA" rod alternative but cannot exclude it.

Salt prevents the formation of clusters: this can be explained by the reduction of the strength of electrostatic interactions. However, surprisingly, salt does not influence strongly the rod-like structures, as far as we can see from our limited set of samples. It is therefore questionable that the interactions are purely electrostatic. H bonds could play an important role. More studies should be done along this line. Seen from another point of view, the lack of salt effect means that the persistence length of the rod is not sensitive to ionic strength, at variance with the one of Hyaluronan itself[40] which shows the predicted behavior of a model semi-rigid polyelectrolyte[43, 45]. The rigidification of the rod could therefore be ensured not only by electrostatic repulsion along the chain but also by the binding of the lysozyme on the HA. Eventually, and not prejudging the exact origin of the attraction, we can propose the simple picture schematized in **Scheme 3**. Note that the distance between two neighbor proteins should NOT be constant, since no correlation corresponding to a privileged distance is observed in the scattering.

**[Scheme 3]**

The dense objects – clusters- detected at high protein content (diphasic samples, second Porod law at low q) have a size too large ( > 60 nm) to be determined accurately. Anyway, they are noticeably larger than the globules in complexes of lysozyme with PSS, which is flexible ($L_{p0} = 1$ nm). Larger globules have already been observed with pectin, which is semiflexible ($L_{p0} = 10$ nm). Thus in term of clusters size, the effect of chain rigidity is similar with pectin and with hyaluronan. On the contrary, there is not such similarity for the rod behavior: for pectin[36], it has not been observed. This may be due to difference in architecture: hyaluronan is strictly linear, with a single repeating unit, while the backbone of pectin is more complex: though some sections of the chains are linear, blocks with dangling parts of different components are present. Also pectin is only partially charged.

Theoretically, exclusively on the basis of electrostatics, Zhang and Schklovskii[35] have predicted both large dense clusters ("condensation") for $[-]/[+]_{intro}$ close to 1, and complexes of small mass for $[-]/[+]_{intro}$ further from 1. In some cases these complexes can be rod-like (chromatin-like structure). They have also predicted a coexistence of extended parts and neutral globular parts corresponding to a situation of "disproportionation". The scattering in some cases, for $[-]/[+]_{intro}$ close to 1, can be interpreted as such a coexistence of two signals, one from clusters, which may well correspond to the condensates, and one from the rod complex. But, from the discussion above, it appears possible that the rods are neutral and coexist with charged species. Finally other origins than electrostatic ones for the rods are possible. More work on different globular charged species, like other proteins or other types of charged spheres could allow us to answer this question.

### V. Conclusion.

Performing several SANS measurements on HA-lysozyme samples allowed us to observe different complex structures depending essentially on the charge ratio. With an excess of HA, a network-like structure appears. For a charge ratio $[-]/[+]_{intro}$ slightly higher than unity for both pHs, an original rod-like structure extending to a size close to the contour length of the HA chain, for low mass chains, and larger than accessible sizes by SANS (60 nm) for large mass chains, is clearly displayed by the SANS measurements. All features of this scattering can be explained by a picture of proteins aligned along one single HA chain (though one rod could in principle contain several HA chains in bundle). For lower charge ratio and low salt concentration, a phase separation occurs and clusters larger than 60 nm appear. Our data do not allow us to decide whether Lysozyme-HA interactions are purely electrostatic, in particular those leading to the rod shape complexes. Therefore, further studies using other proteins are forecast.

**ACKNOWLEDGMENTS**. We thank Institut Laue Langevin and Laboratoire Léon Brillouin for beam time allocation.

**Supporting Information Available. Figure SI.1.** SANS scattering I(q) at pH = 7.4 shifted by a factor 10, for sample of Ha1 (small mass, 10g/L) and lysozyme. **Figure SI.2.** SANS scattering I(q) at pH = 7.4, for sample of HA2 (10g/L) plus lysozyme 3.32g/L, subtracted from the scattering of pure HA solution at same HA concentration, and scattering of pure lysozyme solution at same concentration. **Figure SI.3.** SANS scattering plotted as $q^4 I(q)$ as a function of q for samples of Ha1 (small chains, 10g/L) and lysozyme for charge ratio -/+ = 0.45, 0.68 and 0.9 at pH 4.7, 0.1 M (same data as in upper curves of Figure 2 in the text). This material is available free of charge via the Internet at http://pubs.acs.org.

**NOTE 1**: Using Equation (2) given in the Discussion section,

$$I_0 = \lim_{q \to 0} I(q) = \Phi_{Lyso} . \Delta\rho_{Lyso}^2 . v_{Lyso} \quad (2)$$

and the values of density of contrast $\Delta\rho_{Lyso}^2 = 14.44 \ 10^{20} \ cm^{-4}$ and of volume $15000.10^{-24} \ cm^3$ given in Table 5, we obtain:

$I_0 = \Phi_{Lyso} . 14.44 \ 10^{20} . 15000 \ 10^{-24} = 21.7 \ \Phi_{Lyso}.$

Using $I_0 = 0.28 \ cm^{-1}$, we obtain $\Phi_{Lyso} = 1.29 \ 10^{-2}$ .

Since $\Phi_{Lyso} = C_{Lyso}. V$ (here $C_{Lyso} = 20 \ g/L = 2. \ 10^{-2} \ g/cm^3$) we get a specific volume $V = 1/d$ (*d* density) for lysozyme $0.645 \ g/cm^3$ comparable to the value $0.63 \ g/cm^3$ evaluated formerly as given in **Table 5**.

**NOTE 2**: In the $q^{-1}$ range, the scattering intensity from small mass HA1 complexes quasi overlaps the ones for the large mass (HA2) at same concentration (8g/L, 0.0484 Å$^{-1}$ : 0.084 cm$^{-1}$ large mass and 0.10 cm$^{-1}$ small mass) at same pH and same concentration, or proportionality to concentration is seen (e.g. at q = 0.05 Å$^{-1}$ , I(q) = 0.2 cm$^{-1}$, for 27 g/L, and 0.11cm$^{-1}$ for 10 g/L).

**NOTE 3**: For 10 g/L, one has I(q) = 0.1 cm$^{-1}$ for q = 0.057 Å$^{-1}$ , giving F from Eq. (1). For $C_{Lyso}$ = 20 g/L, $I_0 \sim 0.28$ cm$^{-1}$ (see NOTE 1), giving $I_0 \sim 0.14$ cm$^{-1}$ for $C_{Lyso}$ = 10 g/L. Finally, we get:

$a$ = 0.14 $\pi$. / (0.1 * 0.057) = 78.5 Å.

Another way of evaluating $a$ , more direct, is the following: in insert of **Figure 5**, the scattering from lysozyme - renormalized to 10g/L- and the rod scattering cross each other at q = 0.04 Å$^{-1}$. Using equality between Eq. (1) and Eq. (2), and because the lysozyme scattering is quasi-constant in this "flat" regime, this equality is equivalent to $\pi$ /(q.$a$) = 1, i.e. $a$ = $\pi$ /q $\sim$ 78.5 Å.

**REFERENCES**

(1)  Stern, R.; Asari, A.A.; Sugahara, K.N. *Europ. J. of Cell Biol.* **2006**, *85*, 699-715.

(2)  Li, M.; Rosenfeld, M.; Rolondo, Vilar, E.; Cowman, M.K. *Arch. Biochem. Biophys.* **1997**, *341*, 245-250.

(3)  Kogan, G.; Soltes, L.; Stern, R.; Gemeiner, P. *Biotechnol. Lett.* **2007**, *29,* 17-25.

(4)  Miltner, O.; Schneider, U.; Siebert, C. H.; Niedhart, C.; Uwe Niethard, F. *Osteoarthristis and cartilage*, **2002**, *10*, 680-686.

(5) Moss, J.; Van Damme, M.-P.; Murphy, W.; Preston, B. Arch. *Biochem. Biophys.* **1997**, *348*, 49-55

(6) Cleland, R. L.; Wang, J.; Detweiler D. M. *Macromolecules* **1982**,15, 386-395

(7) Grymonpre, K. Staggemeir, B.; Dubin, P.; Mattison K.W. *Biomacromolecules* **2001** 2, 422-

429.

(8)  Malay, Ö.; Bayrakter, O.; Batigün, A.; *Int. J. of Biol. Macromolecules* **2007**, *40*, 387-393.

(9)  Lenormand, H.; Descherel, B.; Tranchepain, F.; Vincent, J.-C. *Biopolymer* **2008**, *89*, 1088-1103.

(10)  Esquenet, C.; Buhler, E. *Macromolecules* **2002**, *35*, 3708.

(11)  Tribet, C.; Radeva, T. *Surfactant science Series « physical chemistry of polyelectrolytes », Ed. M. Dekker*, New York, **1999**, *19*, 687-741.

(12)  Cooper, C. L.; Dubin, P.L.; Kayitmazer, A.B.; Turksen, S. *Curr. Opin. Coll. Interface Sci.* **2005**,*10,* 52-78.

(13)  Turgeon, S. L.; Schmitt, C.; Sanchez, C. *Curr. Opin. Coll. Interface Sci.*, 2007, 12, 166-178.

(14)  Ball, V.; Winterhalter, M.; Schwinte, P.; Lavalle, Ph.; Voegel J.-C.; Schaaf, P. *J. Phys. Chem. B* **2002**, *106*, 2357-2364

(15)  Gummel, J.; Boué, F.; Demé, B.; Cousin, F. *J. Phys. Chem. B,* **2006**, *110*, 24837-24846.

(16)  Park, J.M.; Muhoberac, B.B.; Dubin, P.L. ; Xia, J. *Macromolecules* **1992**, 25, 290-295.

(17)  Tsuboi, A.; Izumi, T.; Hirata, M.; Xia, J.; Dubin, P. J.; Kokufuta, E. *Langmuir* **1996**, *12*, 6295-6303.

(18)  De Kruif, C.G.; Weinbreck, F.; de Vires, R. *Curr. Opin. Colloid Interface Sci.* **2004**, *9*, 340-349.

(19)  Schmitt, C.; Palma da Silva, T.; Bovay, C., Rami-Shojaei, S., Frossard, P., Kolodziejczyk, E., Leser M.E. *Langmuir* **2005**, *21*, 7786-7795.

(20)  Gummel, J.; Cousin, F.; Boué, F. *J. Am. Chem. Soc.* **2007**, *129* (18), 5806-5807.

(21) Turgeon, S.L.; Schmitt, C.; Sanchez, C. *Curr. op. Coll. Interf. Sc.* **2007**,*12*, 166-178.

(22) De Kruif, C.G.; Tuinier, R. *Food Hydrocolloids,* **2001**, *15*, 555-563.

(23) Girard, M.; Sanchez, C; Laneuville; S.I.; Turgeon; S.; Gauthier, S. *Colloids and Surfaces B*, **2004**, *35*, 15-22.

(24) Gummel, J.; Cousin, F.; Clemens, D.; Boué, F. *Soft Matter* **2008**, *4*, 1653–1664.

(25) Gummel, J.; Cousin, F.; Boué, F. *Macromolecules* **2008**, *41*, 2898-2907.

(26) Seyrek , E.; Dubin, P. L.; Tribet, C.; Gamble, E. A. *Biomacromolecules* **2003**, *4(2)*, 273-282.

(27) Borrega, R.; Tribet, C.; Audebert, R. *Macromolecules* **1999**, *32*, 7798-7806.

(28) Petit, F; Audebert, R; Iliopoulos, I. *Colloid Poly. Sci.* **1995**, *273*, 777-781.

(29) Cousin, F.; Gummel, J.; Ung, D.; Boué, F. *Langmuir* **2005**, *21*, 9675-9688.

(30) Stoll, S.; Chodanowski, P. *Macromolecules* **2002**, *35*, 9556-9562.

(31) Ulrich, S.; Laguecir, A.; Stoll S. *Macromolecules* **2005**, *38*, 8939-8949.

(32) Kayitmazer, A. B.; Shaw, D.; Dubin, P. L. *Macromolecules* **2005**, 38(12), 5198-5204.

(33) Carlsson F., Malmsten M., Linse P. *J. Am. Chem. Soc.* **2003**, *125*, 3140-3149.

(34) Ulrich, S.; Seijo M.; Laguecir, A.; Stoll S. *J. Phys. Chem. B* **2006**, *110*, 20954-20964.

(35) Zhang, R; Shklovskii, B. I. *Phys. Rev. E* **2004**, *69*, 021909.

(36) Schmidt, I.; Cousin, F.; Huchon, C; Boué, F.; Axelos, M. *Biomacromolecules* **2009**, 10 (6) 1346-1357.

(37) Kayitmazer, A.B.; Sabina P. Strand, S. P. ; Tribet, C.; Jaeger, W., Dubin, P.L. *Biomacromolecules* **2007**, *8*, 3568-3577.

(38) Bohidar, H.; Dubin, P. L.; Majhi, P. R.; Tribet, C.; Jaeger, W. *Biomacromolecules* **2005**, *6*, 1573-1585.

(39) Wang, X; Li, Y; Lal, J; Huang, Q. *J. Phys. Chem. B* **2007**, *111*, 515-520.

(40) Santinath Singh, S.; Aswal V.K.; Bohidar H.B. *Int. J. Biological Macromolecules* **2007**, *41*, 301-307.

(41) Hayashi, K.;Tsutsumi, K. ; Nakajima, F.; Norimye, T.; Teramoto, A. *Macromolecules* **1996** *28*, 3824-3830.

(42) Buhler, E. ; Boué, F. *Macromolecules* **2004**, *37* 1600-1610

(43) Bonnet, F; Schweins, R; Boué, F; Buhler, E. *EuroPhysicsLetters* **2008**, *83*, 48002.

(44) www.iue.tuwien.ac.at/phd/windbacher/node63.html

(45) Dubois, E. ; Boué, F. *Macromolecules* **2001**, *34,* 3684-3697

(46) Brûlet, A; Boué, F; Cotton, J.-P. *J. Physique II (Paris)* **1996**, *6*, 885-891.

(47) Odijk, T. *J. Polym. Sci., Polym. Phys.* **1977**, *15*, 477.

(48) Skolnick, J.; Fixman, M. *Macromolecules* **1977**, *10*, 944.

(49) Hampe, O.; Tondo, C.; Hassonvoloch, A. *Biophysical Journal,* **1982** *40*, 77-82

(50) Vidal, O.; Robert, M. C.; Boué, F. *J. Crystal Growth* **1998**, *192*, 271-281.

**TABLES:**

| Concentration HA2 (g/L) | Concentration lysozyme (g/L) | Charge ratio [-]/[+]$_{intro}$ |
|---|---|---|
| 10 | 3.32 | 10.7 |
| 10 | 6.6 | 5.36 |
| 10 | 8 | 4.4 |
| 10 | 10 | 3.6 |
| 10 | 15 | 2.4 |
| 10 | 20 | 2.2 |
| 10 | 26.4 | 1.35 |
| 10 | 40 | 1.1 |

**Table 1.** Composition of the different mixtures lysozyme/HA2 in buffer pH 4.7, I = 0.1 M.

| Concentration HA2 (g/L) | Concentration lysozyme (g/L) | Charge ratio $[-]/[+]_{intro}$ |
|---|---|---|
| 10 | 3.32 | 13.4 |
| 10 | 6.64 | 6.7 |
| 10 | 8 | 5.5 |
| 10 | 10 | 4.5 |
| 10 | 40 | 1.4 |

**Table 2.** Composition of the different mixtures lysozyme/HA2 at pH 7.4, I = 0.03 M.

| Concentration HA1 (g/L) | Concentration lysozyme (g/L) | Charge ratio $[-]/[+]_{intro}$ |
|---|---|---|
| 10 | 3.32 | 13.4 |
| 10 | 8 | 5.5 |
| 10 | 27 | 1.63 |
| 10 | 40 | 1.4 |

**Table 3.** Composition of the different mixtures for lysozyme/HA1 (small molecular weight), in buffer at pH 7.4, I = 0.03 M.

| Instrument | Diaphragm (mm$^2$) | Wavelength (Å) | Distances (m) | Collimations (m) |
|---|---|---|---|---|
| PACE | 7 mm diameter | 6 | 1 | 2.5 |
| | | 6 | 4.5 | 5 |
| | | 13 | 4.5 | 5 |
| D11 | 7 X 10 mm$^2$ | 4.6 | 1.1 | 5.5 |
| | | 4.6 | 5 | 10.8 |
| | | 4.6 | 12 | 13.5 |
| D22 | 7 X 10 mm$^2$ | 6 | 2 | 8 |
| | | 6 | 5 | 8 |
| | | 6 | 17 | 17.5 |

**Table 4.** Configuration of the different SANS experiments.

| Species | $V = 1/d$ | $V_{mol}$ $cm^3$ | $\rho_{specie}$ | Density of contrast with $D_2O$ $\Delta\rho^2$ $(cm^{-4})$ |
|---|---|---|---|---|
| Lysozyme | 0.63 cm$^3$/g | 15000.10$^{-24}$ | 2.5 10$^{10}$ cm$^{-2}$ | 14.44 10$^{20}$ |
| Hyaluronan | 0.59 cm$^3$/g | 398 10$^{-24}$ | 3.6 10$^{10}$ cm$^{-2}$ | 7.3. 10$^{20}$ |

**Table 5:** Scattering density per unit volume, calculated as $\rho = \Sigma_i n\ b_i\ /(\ \Sigma_i\ m_i v_i \times 1.66 \times 10^{-24}\ )$ where $V$ is the inverse of density $d$, $\rho$ represents the scattering length per unit volume, $b_i$ the scattering length, $m_i$ the mass, and $v_i$ the specific volume of the species i, id est lysozyme or $D_2O$ molecule[36] or hyaluronan statistical unit[42,43].

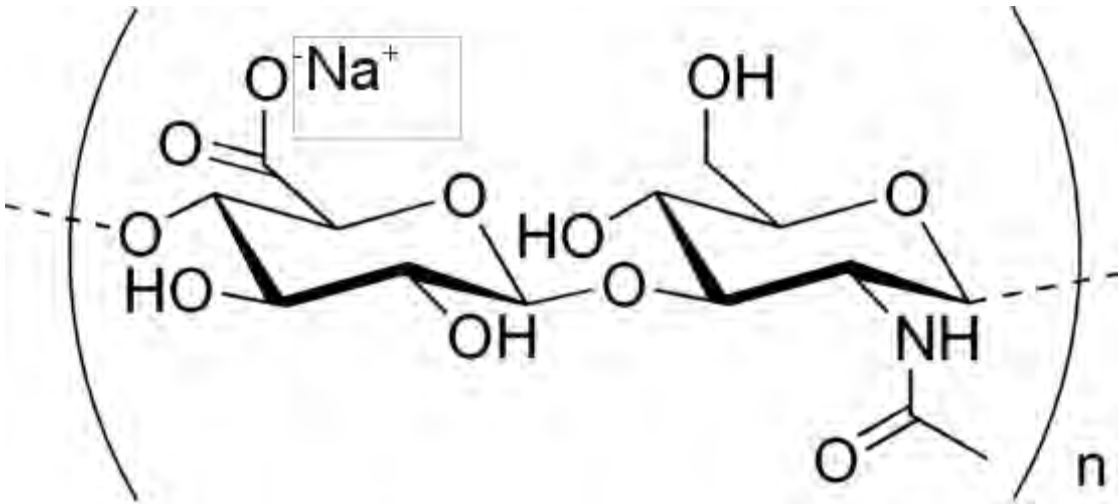

**Scheme 1.** Chemical structure of hyaluronan. The length of the segment is 1 nm.

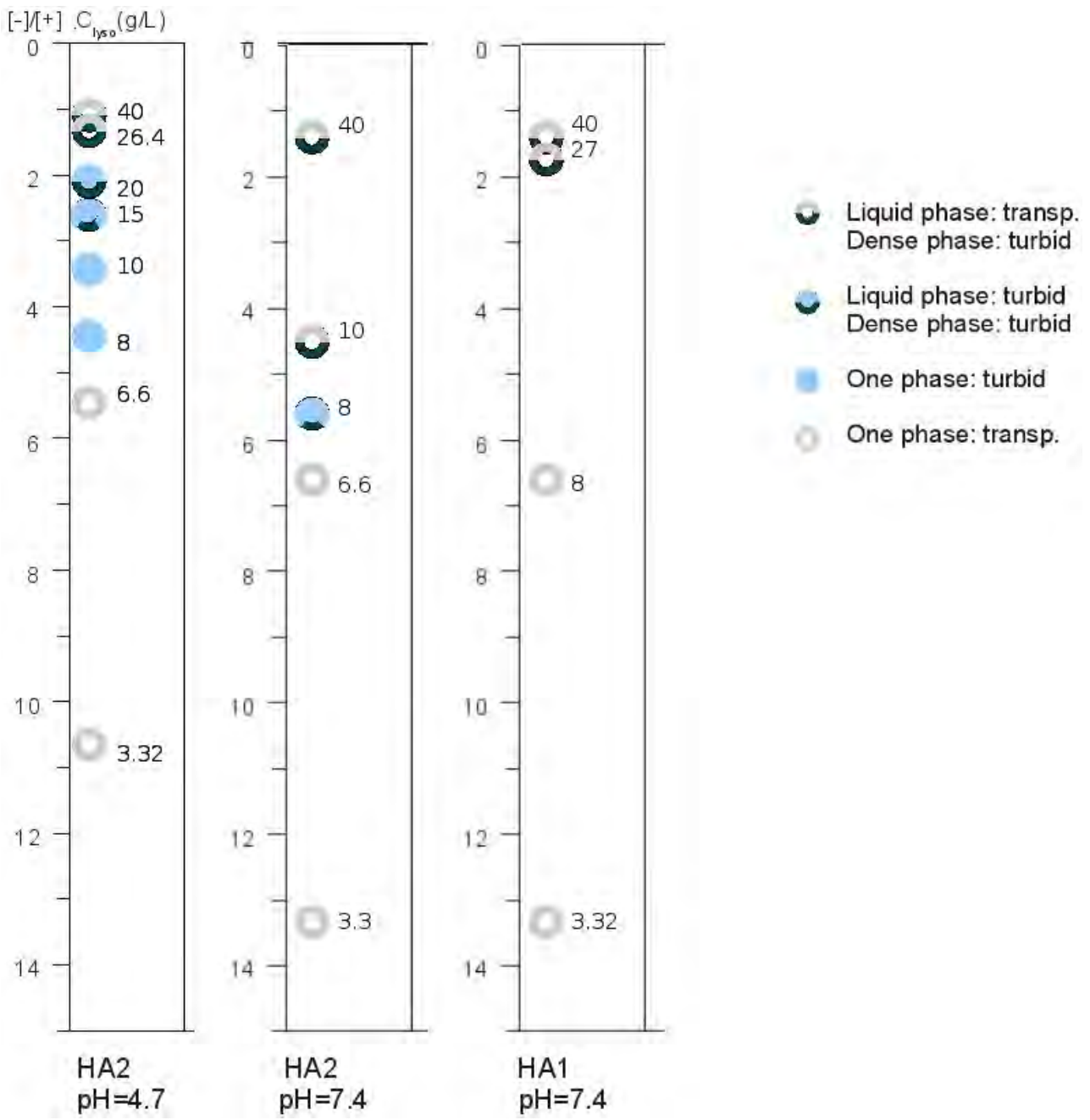

**Scheme 2:** Evolution of the visual aspect of the different samples (HA2 in buffer at pH=4.7, $I_{buffer}$ = 0.1 M and HA2 and HA1 at pH 7.4, , no salt, $I_{buffer}$ = 0.03 M as a function of [-]/[+]$_{intro}$. The HA concentration is kept at 10g/L. The concentration in lysozyme is indicated besides each symbol.

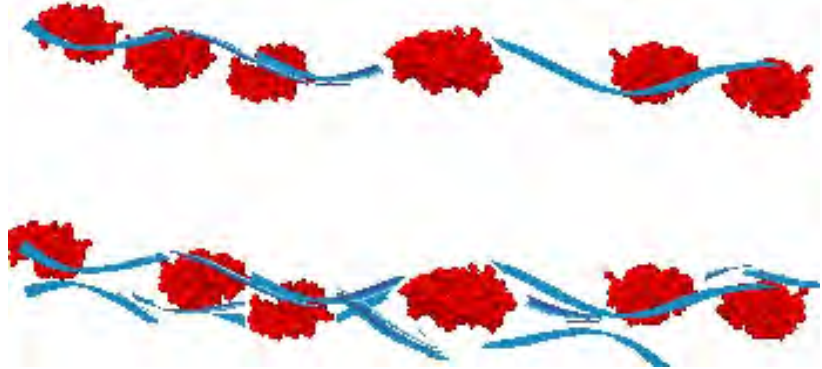

**Scheme 3.** Elementary rodlike complex of hyaluronan and lysozyme. Above single HA chain in the rod. Below: several HA chains per rod, with same protein linear density. The distance between neighbor proteins is not constant.

# FIGURES

**Figure 1.**

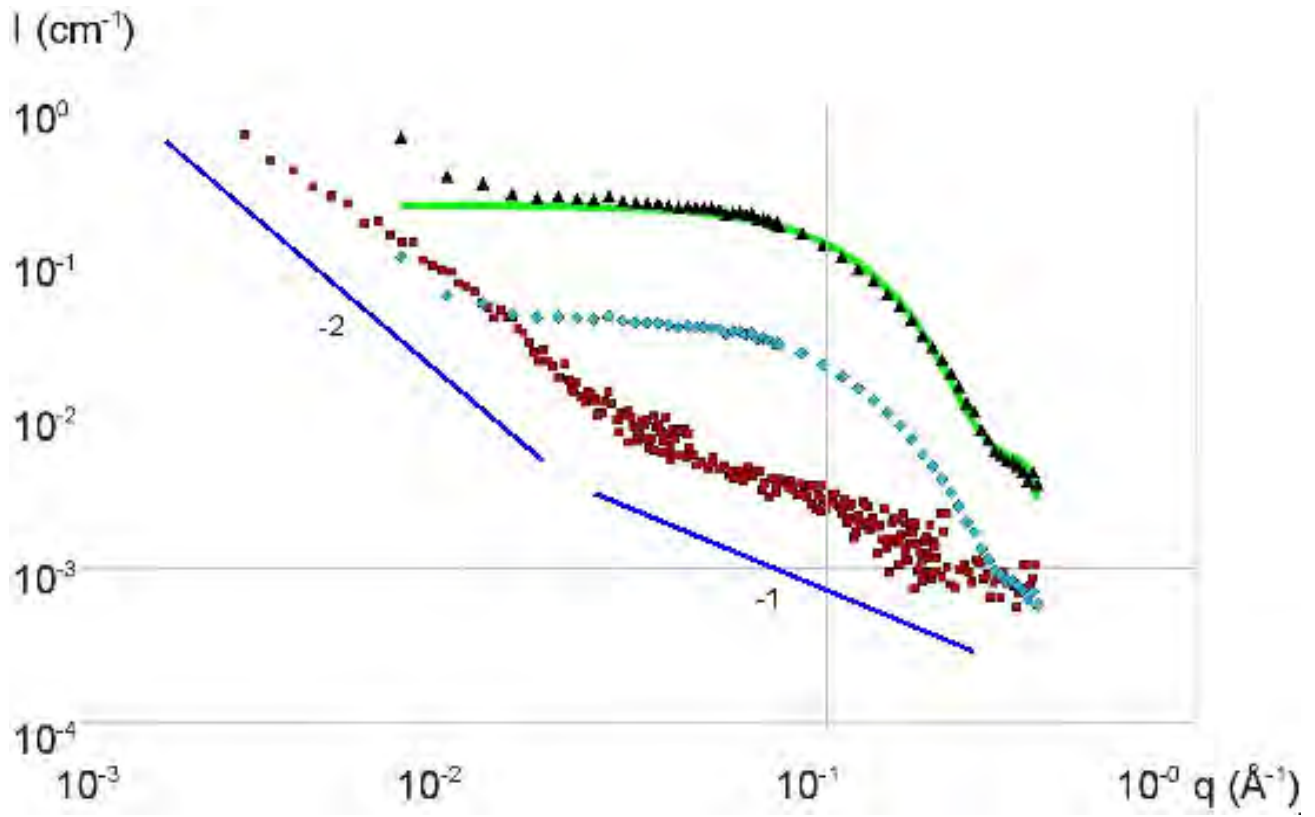

**Figure 1.** SANS I(q) as a function of q for pure HA2 (large mass) solution ( ■ 10g/L,) and pure lysozyme solution (♦ 3.32 g/L and ▲ 20 g/L) in buffer at pH 7.4. In solid line (green), a fit of the lysozyme signal with the form factor of a polydisperse sphere of radius 1.55 nm.

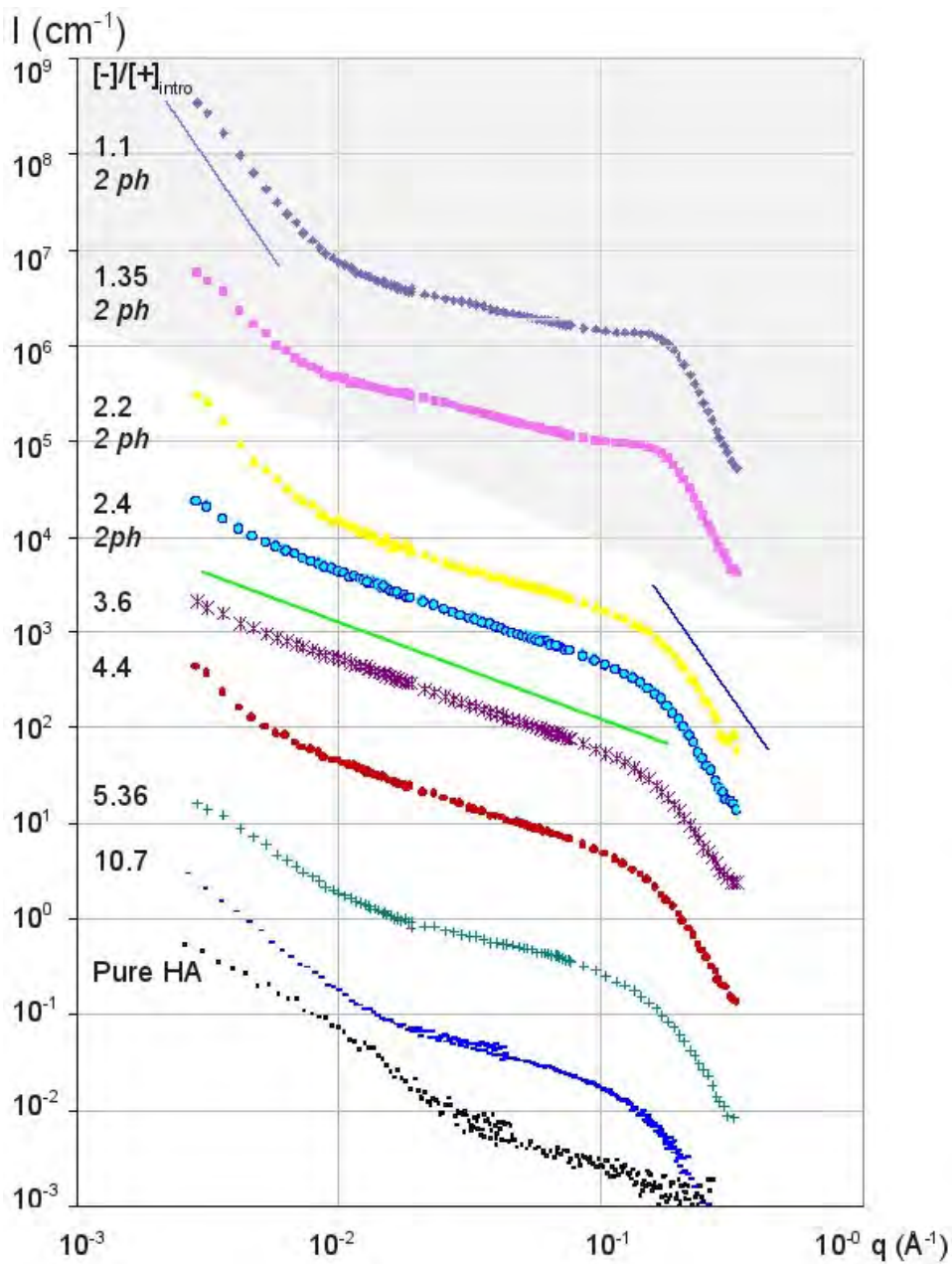

**Figure 2.** SANS I(q) for mixtures of HA2 (large mass, 10 g/L) and lysozyme, at pH = 4.7. On the left of the curves the [-]/[+]$_{intro}$ values and the presence of 2 phases (**2 ph.**) are indicated. [-]/[+]$_{intro}$, decreasing from bottom to top, correspond to increasing lysozyme concentration one phase samples —: 3.32, +: 6.6, ●: 8, *: 10, 2 phases samples ●: 15, ▲: 20, ■: 26 and ◆: 40 g/L. The green and blue straight lines correspond respectively to q$^{-1}$ and q$^{-4}$. Curves are shifted by a factor ten from the one below. The bottom one as well as the scattering of the pure HA2 (–) are in cm$^{-1}$. Data within the grey area come from the dense phases. For 15 g/L, & even more for 20 g/L, a small volume of dense phase lied in the bottom but scattering was measured from the supernatant. Same data normalized by $C_{Lyso}$ (not shown ) show overlap at large q in the region of scattering dominated by lysozyme, except for samples at 15g/L and 20g/L; the latter is too low by a factor 5.

# Figure 3.

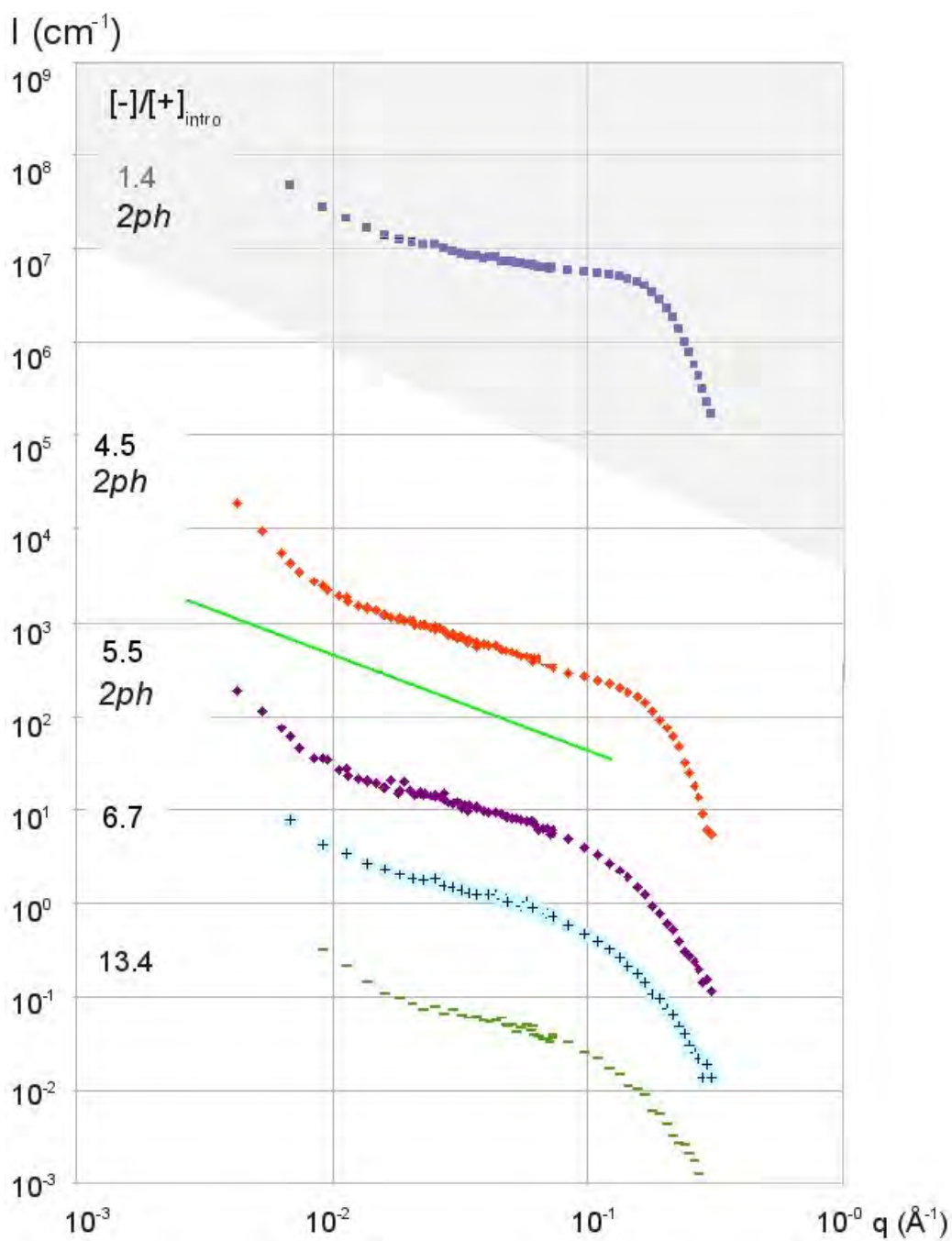

**Figure 3.** SANS I(q) for HA2 (large mass, 10g/L) with lysozyme at pH=7.4 (curves shifted by a factor 10; bottom in cm$^{-1}$). The charge ratio are decreasing from bottom to top, corresponding to increasing lysozyme concentration: one phase samples: —: 3.32, +: 6.6; 2 phases: ♦: 8, ♦: 10, ■: 40 g/L.(Table 2).

**Figure 4.**

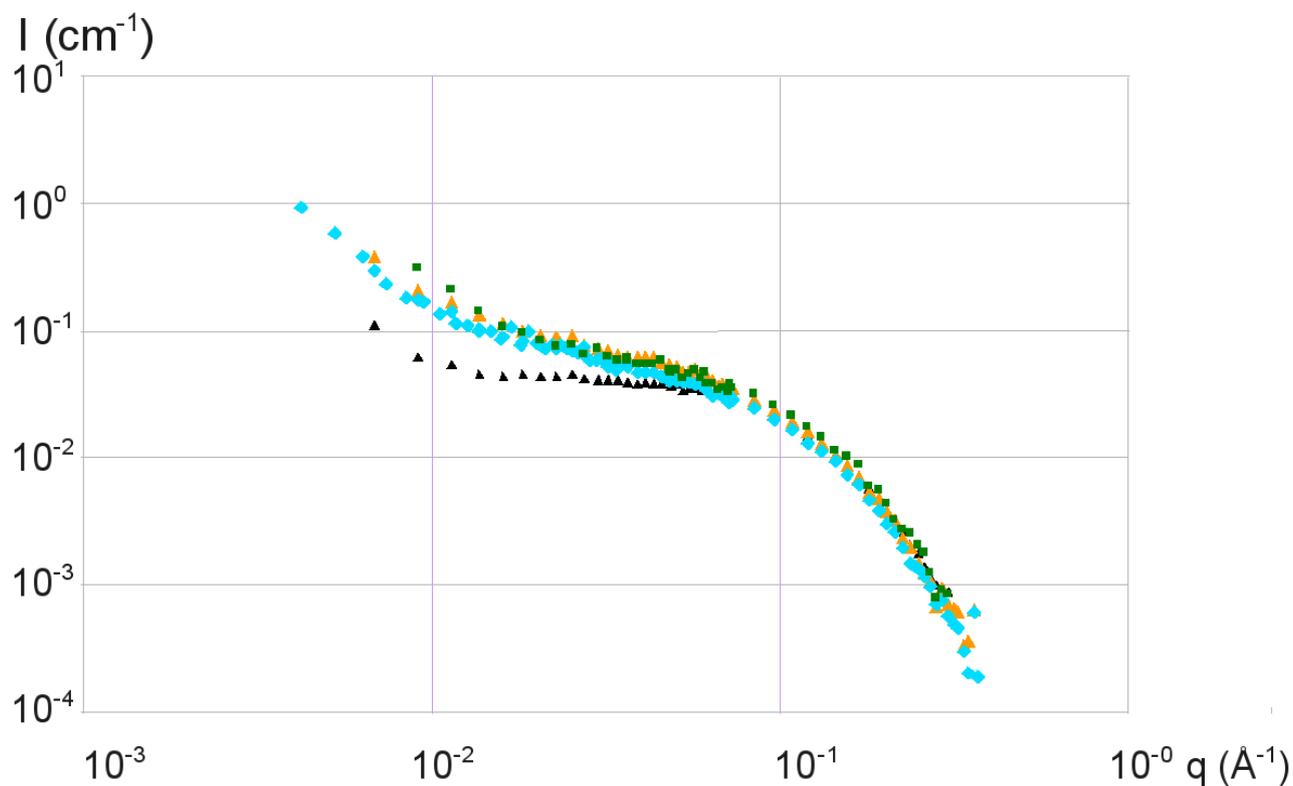

**Figure 4.** SANS scattering I(q) in buffer at pH = 7.4 for a lysozyme solution ( ▲: 3.32g/L) and mixtures of HA2 (large mass, 10g/L) and lysozyme (■:3.32g/L, ◄:6.64g/L, ◆:8g/L (liquid phase), corresponding to the three lowest curves of **Figure 2** and divided - for comparison at same lysozyme concentration- respectively by 1, 2, as expected, and 2 (instead of 2.4 = 8/3.32).

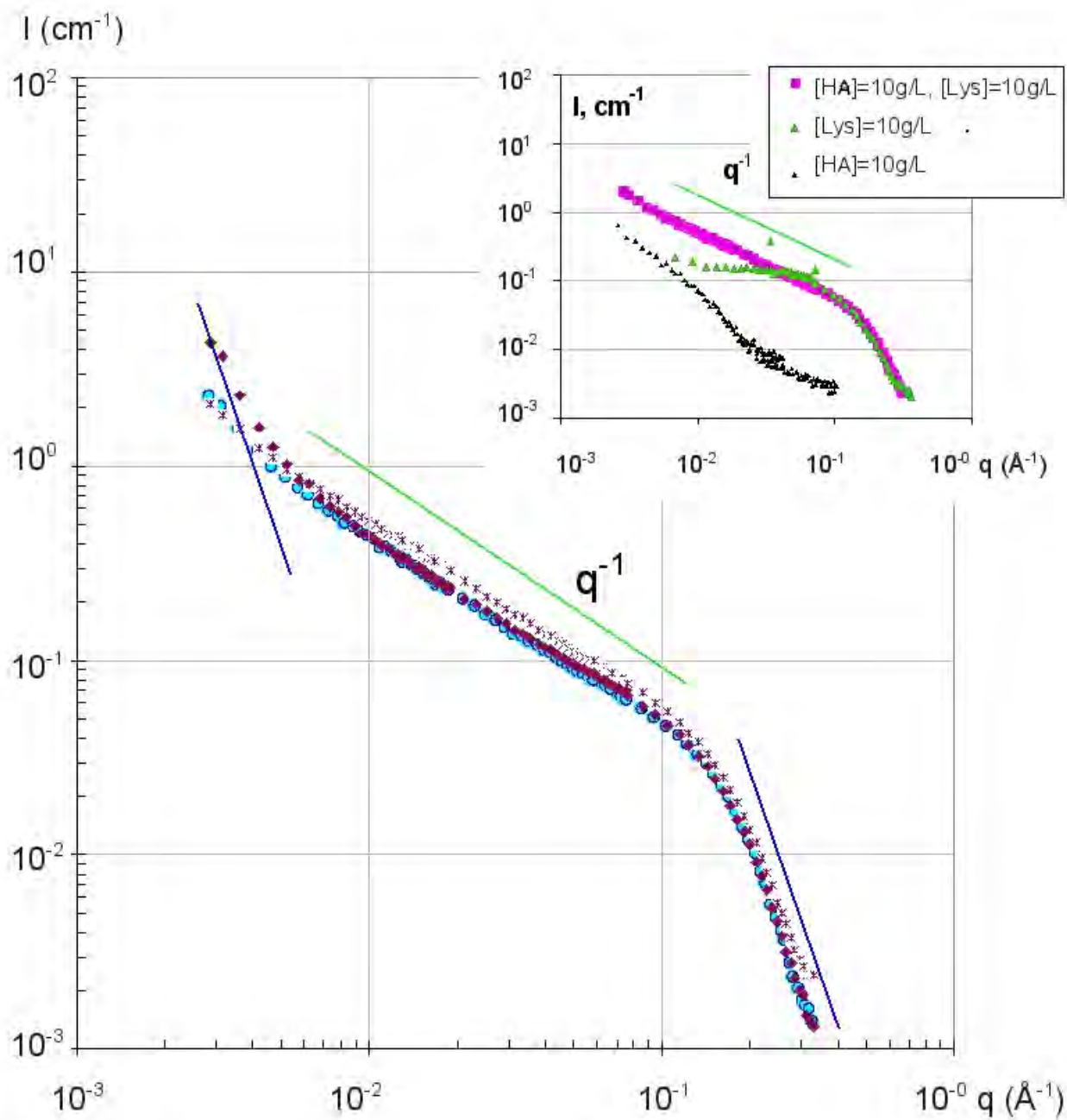

**Figure 5.** SANS I(q) (not normalized by concentration) as a function of q presenting a large q⁻¹ domain for 3 complexes HA2 (large mass, 10g/L), and lysozyme: ♦: 8, *: 10, •: 15 g/L, pH = 4.7 shown already in **Figure 2**. Inset: comparison between HA/lysozyme 10g/L/10g/L, pure lysozyme and pure HA2 at the same concentrations.

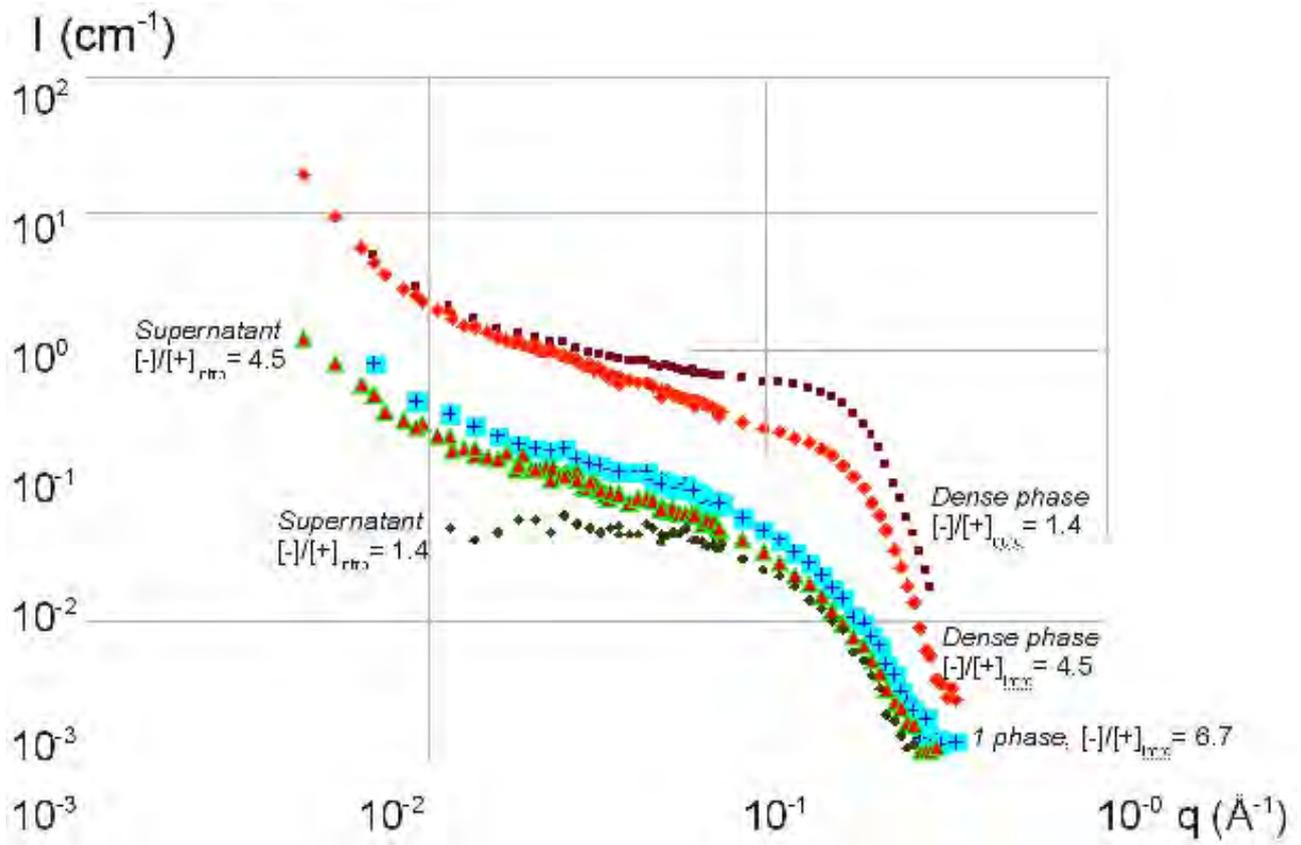

**Figure 6.** Comparison of supernatant and dense phase: SANS scattering I(q) for mixtures of HA2 (large mass, 10g/L) and lysozyme : ▣: 6.6g/L, monophasic ▲; and ◆ are respectively supernatant and dense phase of 10g/L, ◆ and ■ are respectively supernatant and dense phase of 40g/L, at pH=7.4.

**Figure 7.**

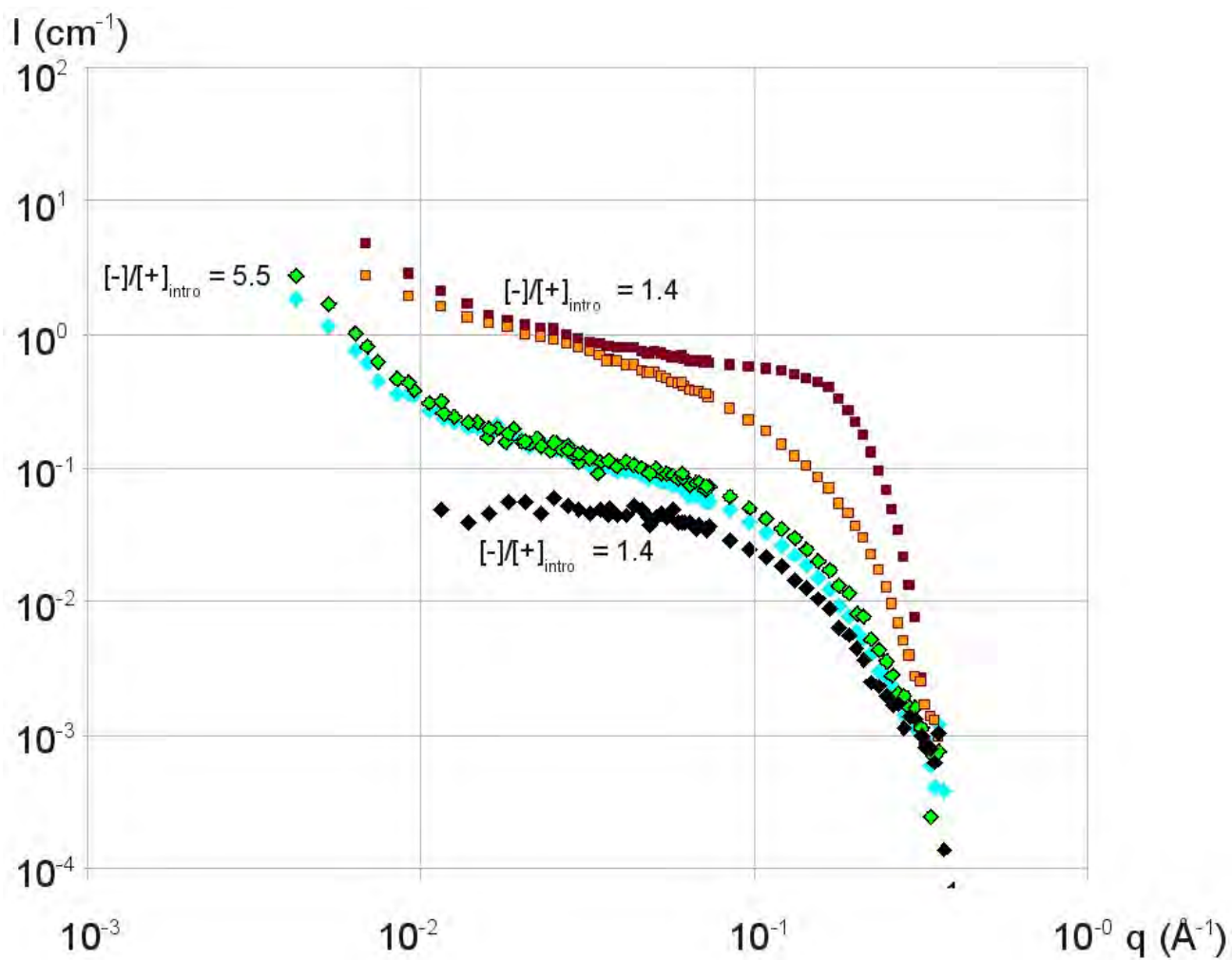

**Figure 7.** Salt effect: SANS I(q) for sample of HA2 (large mass, 10g/L) and lysozyme in buffer solution at pH=7.4. Without salt ($I_{buffer}$ 30 mM): ♦: $C_{Lyso}$ = 8 g/L (one phase), ■: 40 g/L (dense phase), ♦: supernatant of ■. With salt ($I_{buffer}$ 170 mM): ♦: 8 g/L (one phase), ■: 40 g/L (one phase).

**Figure 8.**

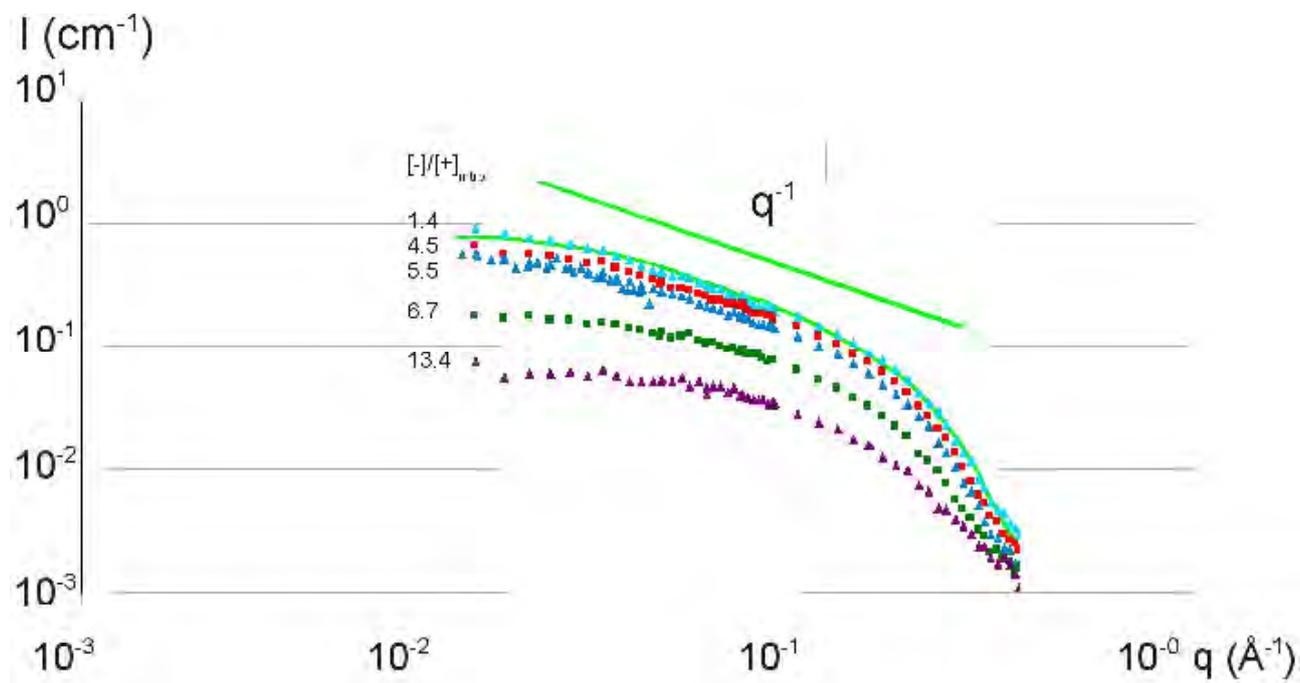

**Figure 8.** SANS scattering as a function of q for sample of HA1 (small mass, 10g/L) and lysozyme:

▲:3g/L, ■:8g/L, ▲:27g/L, ■:40g/L, ▲:40g/L (shaken) in buffer solution at pH = 7.4. In solid line

(green), a fit, for $C_{lyso}$ = 40 g/L, of a rod scattering, length 15 nm , average radius 1.1 nm , volume fraction 0.252, and average contrast $\Delta\rho^2$ = 4.18 $10^{10}$ cm$^{-4}$.

# END of the paper.